\documentclass[12pt,a4paper]{article}
\usepackage[utf8]{inputenc}
\usepackage[english]{babel}
\usepackage[left=1cm,right=1cm,top=2cm,bottom=2cm]{geometry}
\usepackage{array}
\newcolumntype{P}[1]{>{\centering\arraybackslash}p{#1}}
\usepackage{latexsym}
\usepackage{selinput}
\SelectInputMappings{ aacute={á}, ntilde={ñ}}
\usepackage[T1]{fontenc}
\usepackage{textcomp}
\usepackage{amsmath}
\usepackage{amsfonts}
\usepackage{amssymb}
\usepackage{IEEEtrantools}
\usepackage{cases}
\usepackage{caption}
\usepackage[margin=1cm]{caption}
\usepackage[colorlinks=true,linkcolor=blue]{hyperref}
\usepackage{graphics}
\usepackage{graphicx}
\usepackage{verbatim}
\usepackage[toc,page]{appendix}
\providecommand{\keywords}[1]
{
  \small	
  \textbf{\textit{Keywords---}} #1
}
\begin{document}
\title{\textbf{Dual solutions Schrödinger type for Poisson equation in dielectric and magnetic linear media}}
\author{R. Rubiano$^{1}${\footnote{\textit{e-mail: robert.giraldo@uaf.ufcg.edu.br}}}, J. Tapia$^{2}$, and H. González$^{3}$\\
%
$^{1}$Physics postgraduate program at the Science and Technology Center\\
Federal University of Campina Grande-Paraíba\\
$^{2}$Physics graduate program\\
Surcolombiana University-Huila}

\date{}          
\maketitle
\thispagestyle{empty}
\begin{center}
{\small \date{Received: 23/01/2020   / Revised version: 19/05/2020 / Accepted: 15/06/2020}}
\end{center}
\maketitle
\abstract{
Solutions are obtained for the dual form of the Schrödinger equation got from the transformation of Poisson equation for the vector and the scalar potential, in dielectric and magnetic materials, having into account homogeneous isotropic linear mechanisms. We study and apply these dual equation solutions in some specific potentials.} 
\\
\\

\begin{large}
\textbf{Introduction:}
\end{large}
\maketitle
Throughout the first two decades of the twentieth century, the basic conceptions of Quantum Mechanics \cite{De la Pena}, Planck hypothesis and De Broglie suppositions \cite{P. Weinberger} built the structure or the concept of the Non- Relativistic Mechanical Wave Schrödinger \cite{E. Schrodinger}.  Along with the adaptation of the Special Theory of the Relativity, it was built the Relativistic Quantum Mechanics, in which Dirac proved that the electron has an angular momentum at $\frac { \hslash  }{ 2 } $. The relativistic electron from Dirac \cite{A. Dirac} gets represented by spin states $\frac { 1 }{ 2 } $ and $\frac { -1 }{ 2 } $ as well as a positron $\frac { 1 }{ 2 } $ and $\frac { -1 }{ 2 } $. Implicating that we have four quantum numbers which are represented by a spinor. The lookup of connections, or links among the different physics theories and insights has been comprehensive in the last few years \cite{G. Gonzalez}, \cite{V. Rokaj}, \cite{N. Ben Abdallah}. In this article we relate both the wave mechanics of Schrödinger with the electromagnetics theory of Maxwell, including also the vector and scalar potentials for Poisson equation, in specific cases of homogeneous isotropic, lineal dielectric and magnetic medium.

In the present investigation an extensive study is arranged to improve the theoretical knowledge as shown in \cite{G. Gonzalez}, \cite{V. Rokaj}, \cite{G. Gonzalez2}. Based on the González \cite{G. Gonzalez} and Rokaj \cite{V. Rokaj} transformations, it is established a new relationship between electromagnetic theory, and the quantum mechanics, specifically the Poisson equation (homogeneous, linear, isotropic material media) and the Schrödinger equation. In order to convert the equations of electromagnetic into equations of quantum mechanics, due to the wave properties present in matter and in the electromagnetic fields in free space. The most transformations are implemented through mathematical analogies. Following the same methods and procedures in this research, on recent investigations is addressed the simulation problem of relativistic quantum mechanics making use of different physical platforms such as optical structures \cite{M. Mohammad-Ali}, metamaterials \cite{W. Tan}, and ionic traps \cite{L. Lamata}, which provides a way to explore at a macroscopic level many quantum phenomena currently inaccessible at microscopic systems.  
 
Approaches of classical-quantum analogies, investigated so far, due to the duality between matter and optical waves. Most points that prolific results are given by the analogy between optics and quantum phenomena. Since It has not been considered in depth to the present, possible resemblance between these theories may be found in dielectrics or magnetic materials applying some of these transformations. The quantization procedures introduced at early the XX century are not universal. It merits a comprehensive study of other ways of connection between Classical Mechanics and the new contemporary theories of Physics.

The structure of this research is clustered in three main sections: to begin with, in the section \ref{sec:1}, it is set out a review about the contour conditions in the function of the quantum wave and vector and scalar potentials as well. The sec. \ref{sec:2}, introduces a modification to the scalar potential in Poisson equation either in a magnetisable or in a homogeneous isotropic, lineal dielectric medium, that leads in this instances to a Schrödinger dual equation. The sec. \ref{sec:3} aims to obtain solutions to the Schrödinger dual equations found in the previous section to specific problems. As a final point, in the concluding section the conclusions and the analysis of the result are revealed as well.

\section{Relationship between boundary conditions}
\label{sec:1}

Bearing in mind Poisson equation in material medium, which is one of the most significant equations of the electromagnetic theory; this equation correlates mainly the volumetric distribution of the electric charge and the electrostatic potential, that enables to find the electrostatic field if the distribution of the electric charge is well- known. If the behavior either of the materials with homogeneous isotropic linear magnetic permeability or dielectric constant permeability is idealized, the calculations required to determine an electric or magnetic field in a specific material are simplified, since in materials with non-homogeneous dielectric constant, in other words, dependent on the trajectory, would generate more complex connections. Stated otherwise, by using a transformation on the Poisson equation in material medium with either dielectric constants or homogeneous isotropic linear magnetic permeability, it should be obtaining equally important connections or relationships between the Electromagnetic Theory and the Quantum Mechanics regarding  \cite{G. Gonzalez}, \cite{V. Rokaj}.   
\\

Particularly, in this present work, using a simple transformation, it is presented an analogy between an electrostatic, magnetostatic (one-dimensional) problem described by the Poisson equation in matter and the Schrödinger equation. This recent relation allows or facilitates to discover the proper function of the described quantum system, through the use of electrostatic and magnetostatic results for each case. 
\\

In this sense, the boundary conditions to the electric scalar potential and the vector magnetic, in one dimension, in homogeneous isotropic linear materials, can be related to the conditions of a quantum wave. With the following limits \cite{A. Arnold}:

\begin{itemize}
\item[$*$] The electrostatic scalar potential in matter $\Phi \left( \textbf{r} \right) $ is continuous across any limit, excepts for its normal derivate that the surface is discontinuous.

\begin{equation}
\Delta \left( \frac { d\Phi  }{ dn }  \right) =-\frac { { \sigma  }_{ f } }{ \varepsilon  } \label{Eq.1}
\end{equation}
\end{itemize}

\begin{itemize}
\item[$*$] The magnetostatic vector potential $\textbf{A}\left( \textbf{r} \right) $ in matter presents likewise corresponding limits from the electrostatic scalar potential when the magnetostatic vector potential and the surface current are scalar in a space dimension. It is obtained the following equation or expression: 

\begin{equation}
\Delta \left( \frac { \textbf{A} }{ dn }  \right) =-\mu { \textbf{k} }_{ f } \label{Eq.2}
\end{equation}
\end{itemize}

\begin{itemize}
\item[$*$] The quantum proper wave function $\psi \left( z \right) $ is always continuous, except from its derivative which is non-continuous in levels where its potential is infinity $\Phi \left( z \right) ={ V }_{ 0 }\delta \left( z \right) $. It exists a discontinuity in the derivative of the proper wave function in proportion to the wave function in ${ z }_{ 0 }$ (and to the potential force of the delta function). 

\begin{equation}
\left.\frac { d\psi \left( z \right)  }{ dz } \right|_{+\epsilon}-\left.\frac { d\psi \left( z \right)  }{ dz } \right|_{+\epsilon}=constant{ V }_{ 0 }\psi \left( { z }_{ 0 } \right)\label{Eq.3}
\end{equation}

\end{itemize}
 
Due to these limits particularly, it is suggested or stated that the unique way that they can be equivalent is through the existence a potential of delta function. Consequently, the following conversions are proposed \cite{I. Richard}:

\begin{enumerate}
    \item Electrostatic scalar potential \cite{G. Gonzalez}:
    \begin{equation}
    \Phi \left( z \right) ={ V }_{ 0 }\ln { \left( \frac { \psi \left( z \right)  }{ \Lambda   }  \right)  } \label{Eq.4}
    \end{equation}
    \item Magnetostatic vector potential  \cite{V. Rokaj}: 
    \begin{equation}
    { A }_{ x }\left( z \right) ={ A }_{ 0 }\ln { \left( \frac { \psi \left( z \right)  }{ C }  \right)  } \label{Eq.5}
    \end{equation}
\end{enumerate}

In which ${ V }_{ 0 },{ A }_{ 0 },\Lambda  ,C$ are constants that guarantee the dimensional consistency. In this paper, thus, it is suggested using the conversions Eq. \ref{Eq.4} and Eq. \ref{Eq.5} in Poisson equation towards homogeneous isotropic linear material medium, in order to introduce a more general idea about the relationship found in \cite{G. Gonzalez}, \cite{V. Rokaj}. 

\section{Transformation in Poisson's equation that lead to Schrödinger type forms}
\label{sec:2}

By taking Poisson equation for the scalar potential, it is possible to obtain a Schrödinger stationary equation by using a suitable transformation \cite{H. Sadri}. This curious association has enabled us to explore these correspondence practices in order to seek more connections between the electromagnetic theory and the non-relativistic quantum mechanics.
\\

In the first part of this research, we lay out the approaches of the authors \cite{G. Gonzalez}, \cite{V. Rokaj}, taking into account either  homogeneous isotropic linear dielectric  or magnetic medium with similar configuration to elicit Schrödinger stationary equations, also known as dual form of the Schrödinger equation.
\\

The Poisson equation in a homogeneous isotropic dielectric linear medium (see appendix \ref{sec:A}) can be expressed as \cite{A. Stratton}:

\begin{equation}
\frac { { d }^{ 2 }\Phi \left( z \right)  }{ d{ z }^{ 2 } } -{ \chi  }_{ E }\frac { d{ E }_{ z } }{ dz } =-\frac { { \rho  }_{ f } }{ { \varepsilon  }_{ 0 } } \label{Eq.6}
\end{equation}

${ \rho  }_{ f }$ is the volumetric density of free charge  y ${ \chi  }_{ E }$ the dielectric susceptibility. Applying the transformation Eq. \ref{Eq.4} in Eq. \ref{Eq.6} and making appropriate substitutions it is obtained an equation which is mathematically similar to Schrödinger equation:

$$\frac { \varepsilon { a }_{ 0 }^{ 3 }{ E }_{ z }^{ 2 } }{ 2{ \varepsilon  }_{ r } } =\left| \xi  \right| \quad \quad \quad -\frac { { V }_{ 0 }{ a }_{ 0 }^{ 3 }{ \rho  }_{ f }\left( z \right)  }{ 2{ \varepsilon  }_{ r } } =U\left( z \right) \quad \quad { w }_{ d }^{ e }=\frac { \left| \xi  \right|  }{ { a }_{ 0 }^{ 3 } } =\frac { { \varepsilon  }_{ 0 }{ E }_{ z }^{ 2 } }{ 2 } $$

\begin{equation}
-\frac { { \hslash  }^{ 2 } }{ 2m } \frac { { d }^{ 2 }\psi \left( z \right)  }{ d{ z }^{ 2 } } +U\left( z \right) \psi \left( z \right) =-\left| \xi  \right| \psi \left( z \right) \label{Eq.7}
\end{equation}

Where $\left[ { V }_{ 0 }^{ 2 }{ \varepsilon  }_{ 0 }{ a }_{ 0 }^{ 3 } \right] =\left[ \frac { { \hslash  }^{ 2 } }{ m }  \right] $, being  ${ a }_{ 0 }$ an arbitrary  length, ${ { a }_{ 0 } }^{ 3 }$ is a typical volume parameter;  ${ V }_{ 0 }$ a constant potential; ${ \varepsilon  }_{ 0 }$ the electric permittivity constant from the free space. For the case of energy values $\xi <0$ \cite{A. Rabinovitch}, this last equation makes clear that the proper function will be given in terms from the electric potential and the energy density, being $\Lambda $ a normalization constant, and ${ w }_{ d }^{ e }$ the electrostatic energy density:

\begin{equation}
\Psi \left( z \right) =\Lambda { e }^{ \frac { \Phi \left( z \right)  }{ { V }_{ 0 } }  },\quad \quad \xi =-{ w }_{ d }^{ e } \label{Eq.8}
\end{equation}

Similarly, we implemented a transformation on Poisson vector equation in a homogeneous isotropic linear magnetic medium. Something interesting is that in this way we obtain a similar to Eq. \ref{Eq.7}. In order to achieve this association, we take into account again, an extension to the research elaborated by the authors \cite{G. Gonzalez}, \cite{V. Rokaj}, employing in a similar way a transformation for the magnetostatic vector potential. Being the Poisson vector equation (see appendix \ref{sec:B}) in one dimension \cite{A. Stratton}:   

\begin{equation}
\frac { { d }^{ 2 } }{ d{ z }^{ 2 } } { A }_{ x }\left( z \right) =\mu { J }_{ x }\left( z \right) \label{Eq.9}
\end{equation}

${ J }_{ x }\left( z \right) $ is the volumetric density of free current, and  ${ \mu  }={ \mu  }_{ r }{ \mu  }_{ 0 }$ the magnetic permeability of medium, with ${ \mu  }_{ r }$ the relative magnetic permeability. By replacing Eq. \ref{Eq.5} in the equation Eq. \ref{Eq.9} and making the appropriate substitutions as well:

$$\frac { { \mu  }_{ r }{ b }_{ 0 }^{ 3 }{ B }_{ y }^{ 2 } }{ 2\mu  } =\left| \xi  \right| \quad \quad \quad -\frac { A_{ 0 }{ \mu  }_{ r }{ b }_{ 0 }^{ 3 } }{ 2 } { J }_{ x }\left( z \right) =U\left( z \right) \quad \quad { w }_{ d }^{ m }=\frac { \left| \xi  \right|  }{ { b }_{ 0 }^{ 3 } } =\frac { { B }_{ y }^{ 2 } }{ 2{ \mu  }_{ 0 } } \quad \quad $$

\begin{equation}
-\frac { { \hslash  }^{ 2 } }{ 2m } \frac { { d }^{ 2 }\psi \left( z \right)  }{ d{ z }^{ 2 } } +U\left( z \right) \psi \left( z \right) =-\left| \xi  \right| \psi \left( z \right) \label{Eq.10}
\end{equation}

Where $\left[ \frac { { A }_{ 0 }^{ 2 }{ b }_{ 0 }^{ 3 } }{ { \mu  }_{ 0 } }  \right] =\left[ \frac { { \hslash  }^{ 2 } }{ m }  \right] $, being ${ b }_{ 0 }$ an arbitrary  length, such a ${ { b }_{ 0 } }^{ 3 }$is a typical volume parameter; ${ A }_{ 0 }$ a constant potential; ${ \mu  }_{ 0 }$ the magnetic permeability constant of the free space. Analyzing a dual solution for both cases dependent from the potential way, in this case $C$ is the normalization constant and ${ w }_{ d }^{ m }$ the magnetostatic energy density: 

\begin{equation}
\Psi \left( z \right) =C{ e }^{ \frac { A\left( z \right)  }{ { A }_{ 0 } }  },\quad \quad \xi =-{ w }_{ d }^{ m } \label{Eq.11}
\end{equation}

It is important to emphasize that the proper functions Eq. \ref{Eq.8} and Eq. \ref{Eq.10} do not have nodes, thus, it is only possible to determine the fundamental state through this method. This is consistent with the potential uniqueness theorem to which it is guaranteed that the electrostatic and magnetostatic potential, is univocal if the distribution of the electric charge density, and the current density, is specified at all limits.

\section{Magnetostatic and electrostatic solutions to the Schrödinger type equationstions}
\label{sec:3}

Solution methods will be stablished to the eqs. Eq. \ref{Eq.7} and Eq. \ref{Eq.10}, found by mathematic processes based on electrostatic and magnetostatic fields configuration, respectively using Eq. \ref{Eq.8} and Eq. \ref{Eq.11}.

\subsection{Solution based on electrostatic configurations}
\label{sec:3.1}

It is considered a configuration of an odd number of the infinite conductive parallel sheets, spaced out a distance a with a dielectric between them. Where the volumetric charge density $\rho \left( z \right) $ can be expressed as \cite{H. Sadri}: 

\begin{equation}
\rho \left( z \right) =\sigma \sum _{ n=-N }^{ N }{ { \left( -1 \right)  }^{ n } } \delta \left( z-na \right) ,\quad \quad for\quad N=0,1,2... \label{Eq.12}
\end{equation}

Therefore, if we make appropriate substitutions and normalized (see appendix \ref{sec:C}) the function Eq. \ref{Eq.8}, we will found a proper function, that models the wave function corresponding to a quantum particle of mass m moving through the potential energy function, similar to the one-dimensional ionic crystal with a dielectric linear constant \cite{H. Sadri}, to a particular frequency range:

\begin{equation}
\Psi \left( z \right) =\frac { \prod _{ n=-N }^{ N }{ \text{exp}\left( \frac { -{ \alpha  }_{ 1 }m{ \left( -1 \right)  }^{ n+N }\left| z-na \right|  }{ { \hbar  }^{ 2 }{ \varepsilon  }_{ r } }  \right)  }  }{ { \left( { \varepsilon  }_{ r }\frac { { \hslash  }^{ 2 } }{ 2{ \alpha  }_{ 1 }m } \text{exp}\left\{ \frac { -2{ \alpha  }_{ 1 }mNa }{ { \hbar  }^{ 2 }{ \varepsilon  }_{ r } }  \right\} +{ \varepsilon  }_{ r }\frac { { \hslash  }^{ 2 } }{ { \alpha  }_{ 1 }m } N\text{exp}\left( \frac { -{ \alpha  }_{ 1 }m\left( 1+2N \right) a }{ { \hbar  }^{ 2 }{ \varepsilon  }_{ r } }  \right) \sinh { \left( \frac { { a\alpha  }_{ 1 }m }{ { { \varepsilon  }_{ r }\hslash  }^{ 2 } }  \right)  }  \right)  }^{ \frac { 1 }{ 2 }  } }\label{Eq.13}
\end{equation}

Found that the potential energy function is inversely proportional to ${ \varepsilon  }_{ r }$ to the form: 

\begin{equation}
U\left( z \right) =-\frac { { \alpha  }_{ 1 } }{ { \varepsilon  }_{ r } } \sum _{ n=N }^{ N }{ { \left( -1 \right)  }^{ n+N } } \delta \left( z-na \right) \label{Eq.14}
\end{equation}

For $N\ge 1$ and $\xi =-\frac { m{ \alpha  }_{ 1 }^{ 2 } }{ 2{ \hslash  }^{ 2 }{ a }_{ 0 }^{ 3 }{ \varepsilon  }_{ r } } $, in the case of free space ${ \varepsilon  }_{ r }\approx 1$. Where ${ \alpha  }_{ 1 }=\frac { \sigma { V }_{ 0 }{ a }_{ 0 }^{ 3 } }{ 2 } $ is the strength of the potential, $\sigma $  the superficial charge density. This new case presents that is possible to model the form in which a single electron is trapped in a one-dimensional ionic crystal \cite{M. Born} that has a dielectric lineal constant, through proper attractive and repulsive delta functions, produced by the electrostatic configuration of parallel sheets, for negative and positives ions \cite{H. Sadri}.

\subsection{Solution based on magnetostatic configurations}
\label{sec:3.2}

Correspondingly to the previous case, it is considered the problem of an odd number of the infinite parallel sheet \cite{I. E. Irodov}, equally spaced, a distance a with superficial electric current density $\textbf{ k } =\pm { k }_{ 0 }\hat { { e }_{ x } } $, in order to the density of the volumetric current can be written respectively  as  \cite{V. Rokaj}, \cite{H. Sadri}:

\begin{equation}
\textbf{J}=\sum _{ n=-N }^{ N }{ { \left( -1 \right)  }^{ n+N } } { k }_{ 0 }\delta \left( z-nb \right) \hat { { e }_{ x } } ,\quad \quad for\quad N=0,1,2... \label{Eq.15}
\end{equation}

By making appropriate substitutions and normalizing(see appendix \ref{sec:D}) the function Eq. \ref{Eq.11}, it is determined that it is also possible to model a proper wave function that describes the behavior a quantum particle with a m mass that travers a one-dimensional ionic crystal with potential energy Eq. \ref{Eq.16}, which presents a magnetic linear constant:

\begin{equation}
U\left( z \right) =-{ \mu  }_{ r }{ \alpha  }_{ 2 }\sum _{ n=N }^{ N }{ { \left( -1 \right)  }^{ n+N } } \delta \left( z-nb \right) \label{Eq.16}
\end{equation}

\begin{equation}
\Psi \left( z \right) =\frac { \prod _{ n=-N }^{ N }{ \text{exp}\left( \frac { -{ \alpha  }_{ 2 }m{ \mu  }_{ r }{ \left( -1 \right)  }^{ n+N }\left| z-nb \right|  }{ { \hbar  }^{ 2 } }  \right)  }  }{ { \left( \frac { { \hslash  }^{ 2 } }{ 2{ \mu  }_{ r }{ \alpha  }_{ 2 }m } \text{exp}\left\{ \frac { -2{ \mu  }_{ r }{ \alpha  }_{ 2 }mNb }{ { \hbar  }^{ 2 } }  \right\} +\frac { { \hslash  }^{ 2 } }{ { { \mu  }_{ r }\alpha  }_{ 2 }m } N\text{exp}\left( \frac { -{ \alpha  }_{ 2 }{ \mu  }_{ r }m\left( 1+2N \right) b }{ { \hbar  }^{ 2 } }  \right) \sinh { \left( \frac { { b{ \mu  }_{ r }\alpha  }_{ 2 }m }{ { \hslash  }^{ 2 } }  \right)  }  \right)  }^{ \frac { 1 }{ 2 }  } } \label{Eq.17}
\end{equation}

For $N\ge 1$ y $\xi =-\frac { m{ \alpha  }_{ 2 }^{ 2 }{ { \mu  }_{ r } }^{ 2 } }{ 2{ \hslash  }^{ 2 }{ b }_{ 0 }^{ 3 } } $, in the free space case ${ \mu  }_{ r }\approx 1$. Where ${ \alpha  }_{ 2 }=\frac { { A }_{ 0 }{ b }_{ 0 }^{ 3 }{ k }_{ 0 } }{ 2 } $ is the strength of the potential,  ${ k }_{ 0 }$ the superficial current density. This result is a natural consequence of the duality with the electrostatic problem. Thus, the proper function Eq. \ref{Eq.17} that models a wave functions is valid for paramagnetic materials where ${ \chi  }_{ m }$ is positive, as well as for diamagnetic materials where ${ \chi  }_{ m }$ is negative \cite{A. Sepulveda}.
\\

To this extend, the contributions of the opposite charged sheets are canceled by pairs and only the contribution of a single charge sheet \cite{F.J. Dyson}; hence, the expectation value of the potential energy is independent of $N$, in the electrostatic case ${ \left< U \right>  }_{ e }$, as well as in the magnetostatic ${ \left< U \right>  }_{ m }$:

\begin{equation}
{ \left< U \right>  }_{ e }=-\frac { m{ \mu  }_{ r }{ \alpha  }_{ 2 }^{ 2 } }{ { \hslash  }^{ 2 } } \quad \quad \quad \quad { \left< U \right>  }_{ m }=-\frac { m{ \alpha  }_{ 2 }^{ 2 } }{ { \hslash  }^{ 2 }\varepsilon _{ r } } \label{Eq.18}
\end{equation}

\section{Analysis and discussion of results}
\label{sec:4}

The developed proper functions Eq. \ref{Eq.13} and Eq. \ref{Eq.17} will be tested through an in-depth study of the transformations used in  \cite{G. Gonzalez} and \cite{V. Rokaj}, to model analog quantum systems with electrostatic or magnetostatic configurations in matter. Finally, both are evaluated in different material media taking into account constants of magnetic permeability and electrical permittivity. As a complimentary application it was used Wolfram Mathematica 10 for calculations and graph generation.

The proper function Eq. \ref{Eq.13} that models the distribution of uniform electrostatic fields in matter is initially simulated for Nitrogen,

\begin{figure}[hbtp]
\centering
\includegraphics[scale=1]{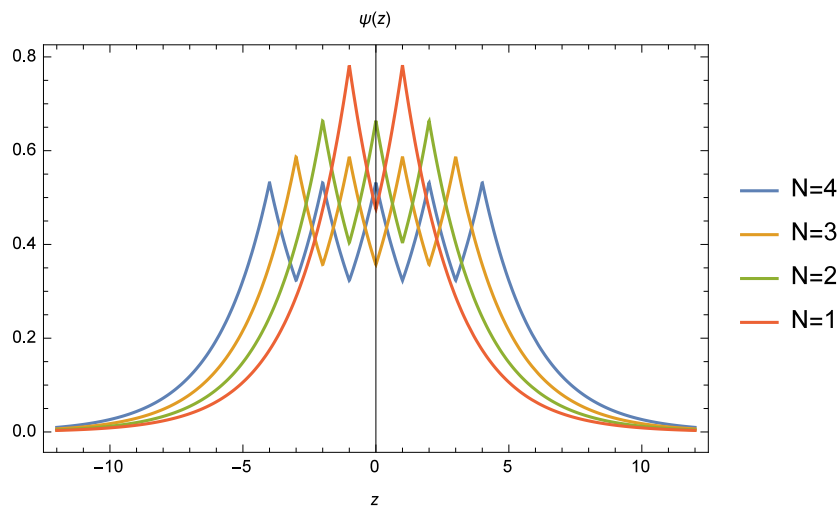}
\caption{Proper function graphic (12) in an only representation for $\varepsilon _{ r }=1.000548$ with different values of  $N$.}
\label{Fig:1}
\end{figure}

In the fig. \ref{Fig:1}. It is shown a descriptive behavior of a triangular potential when using a Dirac delta potential well barrier with Dirichlet boundary conditions, this pattern is corresponding to that presented in one-dimensional ion crystals \cite{H. Sadri}.  For the purpose of comparing the behavior described by the proper function Eq. \ref{Eq.13} with the González's proper function \cite{G. Gonzalez}, it is tested again with a dielectric constant equal to that presented by the free space (${ \varepsilon  }_{ r }\approx 1$). As a result, for each value of $N$ it is found that the pattern of the proper functions varies widely. In subsequent publications by González and Rokaj \cite{G. Gonzalez}, \cite{V. Rokaj}, \cite{G. Gonzalez2} corrections are made to the proper function in free space; however, the relevant graphs are not shown neither it is completely corrected due to errors in the mathematical analysis.

Next, it is performed essays with several dielectric medias with relative electric permittivity constants that differs significantly from Nitrogen (see table \ref{tab:1}), in order to do new evaluations of the proper function Eq. \ref{Eq.13}.

\begin{table}[hbtp]
\centering
\caption{Dielectric Constants}
\label{tab:1}       
\begin{tabular}{P{7cm} P{7cm}}
\hline\noalign{\smallskip}
Material & Relative electric permittivity, ${ \varepsilon  }_{ r }$   \\
\noalign{\smallskip}\hline\noalign{\smallskip}
Nitrogen &  1.000548 \\
Benzene &  2.280000  \\
Salt &  5.900000 \\
Silicone &  11.700000 \\
\noalign{\smallskip}\hline
\end{tabular}
 \caption*{{\small \textit{Note: (unless it is otherwise specified, the values given for 1 atm, 20°C obtained from  Handbook of Chemistry and Physics, 91st ed. (Boca Raton: CRC Press, 2010) \cite{J. D. Griffiths}}}}
\label{Tab:1}
\end{table}


\begin{figure}[hbtp]
\centering
\includegraphics[scale=1]{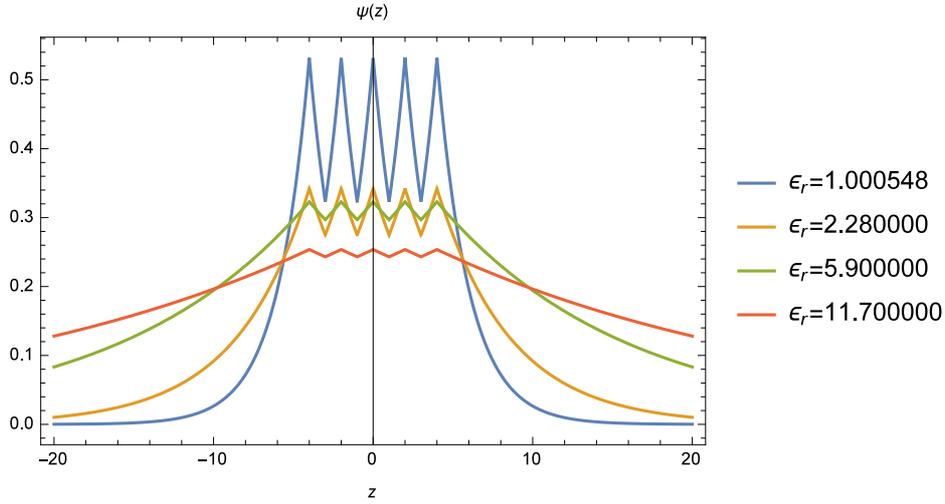}
\caption{Proper function configurations (12) for the different materials shown in table \ref{Tab:1} in $z$ function with $N=4$}
\label{Fig:2}
\end{figure}

In fig. \ref{Fig:2}. Highlights that by increasing the relative electrical permittivity of the material under combined boundary conditions, the proper function extends more smoothly (Neumann boundary condition) and is more slowly variable (Dirichlet boundary condition). The modification of the material between the plates parallel conductors plays a key role in the results. Similarly occurs as in one-dimensional ionic crystals \cite{M. Born} with a designated frequency \cite{F. C. Brown}.
\\ 

\begin{figure}[hbtp]
\centering
\includegraphics[scale=1]{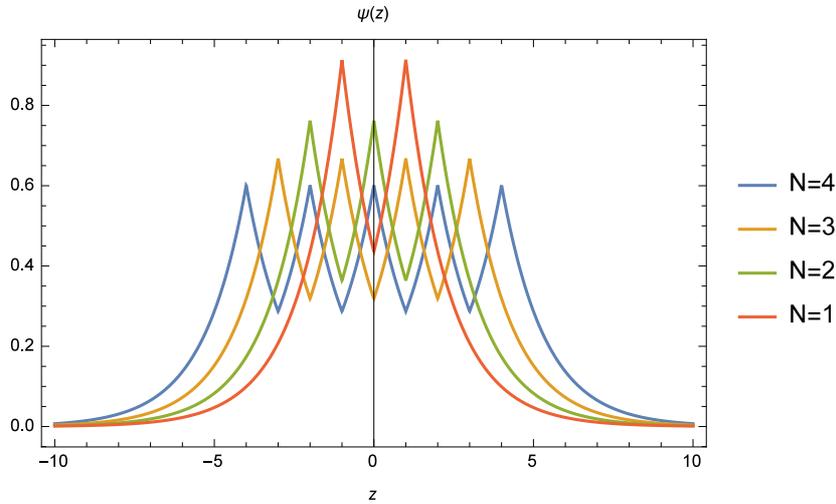}
\caption{Graphics of the proper function (16) in single representation for the Gd with the different values of $N$, with a relative magnetic permeability constant  $\mu _{ r }=1.48$.}
\label{Fig:3}
\end{figure}

The proper function Eq. \ref{Eq.17} models a distribution of uniform magnetostatic fields in matter. It is simulated with Gadolinium (Gd), since it remains linear despite its high magnetization. In fig. \ref{Fig:3}. It is observed how the Gd under Neumann boundary conditions reinforces the magnetostatic field, causing the proper function Eq. \ref{Eq.17} to reach higher values for different set quantities of parallel plates; its expression is influenced by potential energy Eq.\ref{Eq.16}, analog as in one-dimensional ion crystal, contrary in the  case of dielectric materials.

\section{Conclusions}
\label{sec:5}

The present extensive study has been developed through an analytical process in such a way that the solutions have been found accurately; being finally simulated the different proper solutions Eq. \ref{Eq.13} and Eq. \ref{Eq.17}, to complement it, where the constants of electric permeability and magnetic permeability from different medium are also taken into account. In contrast to the research, proposed by González and Rokai for electrostatic and magnetostatic fields configurations in free space (when $\varepsilon _{ r }=1$ y $\mu _{ r }=1$), which does not depend on these constants and shows mathematical errors. This is an advantage for our study since it can be applied in different material mediums either with different electric or magnetic permeabilities, such that the proper function presents a quantum simile or affinity. 
In this extensive study presented it is demonstrated that the proper function Eq. \ref{Eq.13} at high values of electrical permittivity ($1.000548\ll { \varepsilon  }_{ r }$)  stops modeling a quantum system. In contrast to the proper function Eq. \ref{Eq.17} the higher magnetic permeability ($1.48\ll { \mu  }_{ r }$) it acquires much more similarities with a quantum system; it is essential to clarify that these behaviors are characteristic of homogeneous isotropic linear materials by using different boundary conditions. 
At the sight of this work it is concluded that the study conducted based on the research \cite{G. Gonzalez}, \cite{V. Rokaj} to determine the versatility of the transformations on the one-dimensional equation from Poisson in different homogeneous isotropic material medium, has shown its effectiveness stablishing a range of magnetic permeability and electric permittivity for which it is possible an analogy between the electromagnetic theory and the quantum mechanics. It is important to highlight or emphasize that in order to give an explanation to phenomena like the paramagnetism or ferromagnetism it is necessary the quantum theory; therefore, the study of analogies between the classic theory and the quantum theory leads to a better understanding of these phenomena.

\section*{Acknowledgement}
To Surcolombiana University through the Vice-Rector of Research and Social Projection, for the economic support given for the development of this article through the research project 2334. To my friend Ernesto Legra for contributing to its edition.

%
%
%

\begin{appendices}
\section{Deduction of Poisson's equation in an isotropic homogeneous linear dielectric medium.}
\label{sec:A}
In a material medium we must add to the free charge density  ${ \rho  }_{ f }$ the bound charge density  ${ \rho  }_{ b }$:

\renewcommand{\theequation}{\thesection.\arabic{equation}}
\begin{equation}
\nabla \cdot \textbf{E}=\frac { { \rho  }_{ t } }{ { \varepsilon  }_{ 0 } } =\frac { \left( { \rho  }_{ f }+{ \rho  }_{ b } \right)  }{ { \varepsilon  }_{ 0 } } \label{A.19}
\end{equation}

For linear, isotropic dielectrics, the volumetric density of the bound charge is defined as minus the polarization flux in the medium:

\renewcommand{\theequation}{\thesection.\arabic{equation}}
\begin{equation}
{ \rho  }_{ b }=- \nabla   \cdot \textbf{ P } \label{A.20}
\end{equation}

It is known that  ${ \chi  }_{ ij }=0$ si $i=j$, ${ \chi  }_{ xx }={ \chi  }_{ yy }={ \chi  }_{ zz }={ \chi  }_{ e }$, the electrical susceptibility does not depend on the direction:

\renewcommand{\theequation}{\thesection.\arabic{equation}}
\begin{equation}
{ P }_{ x }={ \varepsilon  }_{ 0 }{ \chi  }_{ e }{ E }_{ x },\quad \quad { P }_{ y }={ \varepsilon  }_{ 0 }{ \chi  }_{ e }{ E }_{ y },\quad \quad { P }_{ z }={ \varepsilon  }_{ 0 }{ \chi  }_{ e }{ E }_{ z } \label{A.21}
\end{equation}

Being  $\textbf{ P } $ the electrical polarization, which depends on the electrical susceptibility  ${ \chi  }_{ e }$ of the different materials. In linear, isotropic dielectric media:

\renewcommand{\theequation}{\thesection.\arabic{equation}}
\begin{equation}
\textbf{ P } ={ \varepsilon  }_{ 0 }{ \chi  }_{ e }\textbf{ E } \label{A.22}
\end{equation}

Replacing (A.21) in the (A.19):

 \renewcommand{\theequation}{\thesection.\arabic{equation}}
\begin{equation}
 { \rho  }_{ b }=-{ \varepsilon  }_{ 0 }{ \chi  }_{ e } \nabla  \cdot \textbf { E } \label{A.23}
\end{equation}

Starting from $\textbf { E } =- \nabla   \Phi \left( \textbf{ r }  \right) $, and adding the term of interaction with matter:

\renewcommand{\theequation}{\thesection.\arabic{equation}}
\begin{equation}
{ \nabla  }^{ 2 }\Phi \left( \textbf { r }  \right) =-\left( \frac { { \rho  }_{ f }+{ \rho  }_{ b } }{ { \varepsilon  }_{ 0 } }  \right) \label{A.24}
\end{equation}

Thus, must be added the bound charge density \ref{A.23} in \ref{A.24} to get Poisson's equation in a homogeneous and isotropic dielectric.

\renewcommand{\theequation}{\thesection.\arabic{equation}}
\begin{equation}
{ \nabla  }^{ 2 }\Phi \left( \textbf{ r }  \right) =\frac { { \varepsilon  }_{ 0 }{ \chi  }_{ E } \nabla   \cdot \textbf { E } -{ \rho  }_{ f } }{ { \varepsilon  }_{ 0 } } \label{A.25}
\end{equation}

\section{Deduction of the Vector Poisson equation in one dimension}
\label{sec:B}

Understanding that a magnetizable medium has a magnetization current ${ \textbf{ J }  }_{ M }$, and considering the presence of free currents ${ \textbf { J }  }_{ f }$, the rotor of the magnetic field can be written as:

\renewcommand{\theequation}{\thesection.\arabic{equation}}
\begin{equation}
 \nabla   \times \textbf { B } ={ \mu  }_{ 0 }\left( { \textbf{ J }  }_{ M }+{ \textbf { J }  }_{ f } \right) \label{B.26}
\end{equation}

For a homogeneous linear magnetizable medium ${ \textbf{ J }  }_{ M }$, the magnetization current represents the magnetization circulation density $\textbf { M } $, which is the term that provides information about the interaction of the magnetic field with matter:

\renewcommand{\theequation}{\thesection.\arabic{equation}}
\begin{equation}
{ \textbf{ J }  }_{ M }= \nabla   \times \textbf { M } \label{B.27}
\end{equation}

By rewriting the rotor of the magnetic field \ref{B.26} using the \ref{B.27}:

\renewcommand{\theequation}{\thesection.\arabic{equation}}
\begin{equation}
 \nabla   \times \left( \frac { \textbf{ B }  }{ { \mu  }_{ 0 } } -\textbf { M }  \right) ={ \textbf { J }  }_{ f } \label{B.28}
\end{equation}

The intensity vector of the magnetic field is defined as $\textbf { H } =\frac { \textbf{ B }  }{ { \mu  }_{ 0 } } -\textbf { M } $, finding the modified Ampère's law:

\renewcommand{\theequation}{\thesection.\arabic{equation}}
\begin{equation}
 \nabla  \times \textbf{ H } = { \textbf{ J }_{ f } } \label{B.29}
\end{equation}

In homogeneous and isotropic linear materials it is established that the magnetization vector is proportional to the intensity of the magnetic field:

\renewcommand{\theequation}{\thesection.\arabic{equation}}
\begin{equation}
\textbf{ M } ={ \chi  }_{ m }\textbf{ H } \label{B.30}
\end{equation}

Consequently, it can be rewritten  $\textbf { H } $ as follows:

\renewcommand{\theequation}{\thesection.\arabic{equation}}
\begin{equation}
\textbf { H } =\frac { \textbf{ B }  }{ { \mu  }_{ 0 } } -{ \chi  }_{ m }\textbf { H } \label{B.31}
\end{equation}

As a result it is found that the magnetic fields are times the intensity of the magnetostatic field:

\renewcommand{\theequation}{\thesection.\arabic{equation}}
\begin{equation}
\textbf { B } =\mu \textbf{ H } \label{B.32}
\end{equation}

As the magnetic permeability depends on the magnetic susceptibility of the material, the permeability of the material medium must be defined as:

\renewcommand{\theequation}{\thesection.\arabic{equation}}
\begin{equation}
\mu ={ \mu  }_{ 0 }\left( 1+{ \chi  }_{ m } \right) \label{B.33}
\end{equation}

Where ${ \chi  }_{ m }$ it is known as magnetic susceptibility. For diamagnetic materials  ${ \chi  }_{ m }$ it is negative, such that $\mu <1$; while in paramagnetic materials ${ \chi  }_{ m }$ it is positive, such that $\mu >1$. This relationship becomes more complex in non-homogeneous, non- isotropic materials, because magnetic susceptibility tends to become a tensor ${ \chi  }_{ m }\rightarrow { \hat { \chi  }  }_{ m }$ where its value depends on the position in the material. Made this exception, and introducing  \ref{B.32} and \ref{B.33} in \ref{B.30}, it is found that:

\renewcommand{\theequation}{\thesection.\arabic{equation}}
\begin{equation}
 \nabla   \times \left( \frac { \textbf { B }  }{ { \mu  }_{ 0 } } -{ \chi  }_{ m } \textbf{ H }  \right) = \nabla   \times \left( \frac { \textbf { B }  }{ { \mu  }_{ 0 } } -{ \chi  }_{ m }\frac { \textbf{ B }  }{ { \mu  } }  \right) ={ \textbf { J }  }_{ f } \label{B.34}
\end{equation}

\renewcommand{\theequation}{\thesection.\arabic{equation}}
\begin{equation}
 \nabla  \times \textbf { B } ={ \mu \textbf{ J }  }_{ f } \label{B.35}
\end{equation}

Recalling that the magnetic field can be expressed as the rotational of the magnetic vector potential

\renewcommand{\theequation}{\thesection.\arabic{equation}}
\begin{equation}
 \nabla   \times \left(  \nabla   \times \textbf{ A }  \right) = \nabla   \cdot \left(  \nabla   \cdot \textbf { A }  \right) -{ \nabla  }^{ 2 } \textbf{ A } =\mu { \textbf{ J }  }_{ f } \label{B.36}
\end{equation}

In the stationary margin the equation is reduced to:

\renewcommand{\theequation}{\thesection.\arabic{equation}}
\begin{equation}
{ \nabla  }^{ 2 } \textbf{ A } =-\mu { \textbf{ J }  }_{ f } \label{B.37}
\end{equation}

Where finally \ref{B.37} turns out to be the modified Poisson vector equation for homogeneous isotropic materials.

\section{Derivation of the eigenfunction (12)}
\label{sec:C}

Considering the general case $N\ge 1$, $\sigma =\sum _{ n=-N }^{ N }{ { \left( -1 \right)  }^{ n } } \sigma $, the equation describing an electrostatic potential with a configuration of conductive, infinite, odd, spaced an arbitrary distance plates ${ z }_{ n }=na$ is established as:  

\renewcommand{\theequation}{\thesection.\arabic{equation}}
\begin{equation}
\Phi \left( z \right) =\frac { -\sigma  }{ 2\varepsilon  } \sum _{ n=-N }^{ N }{ { \left( -1 \right)  }^{ n } } \left| z-na \right| \quad \quad \label{C.38}
\end{equation}

Substituting \ref{C.38} in Eq. \ref{Eq.8}:

\renewcommand{\theequation}{\thesection.\arabic{equation}}
\begin{equation}
\Psi \left( z \right) =\Lambda \text{exp}\left\{ \frac { -\sigma  }{ 2{ V }_{ 0 }\varepsilon  } \sum _{ n=-N }^{ N }{ { \left( -1 \right)  }^{ n } } \left| z-na \right|  \right\} \label{C.39}
\end{equation}

Using dimensional analysis on the constants that accompany the exponential $\left[ \frac { \sigma  }{ { V }_{ 0 }{ \varepsilon  }_{ 0 } }  \right] =\left[ \frac { 1 }{ l }  \right] $, an eigenfunction that describes the behavior of the electrostatic potential in the distribution of parallel plates with a homogeneous dielectric inside is determined, as follows:

\renewcommand{\theequation}{\thesection.\arabic{equation}}
\begin{equation}
 \Psi \left( z \right) =\Lambda \text{exp}\left\{ \frac { -1 }{ 2l{ \varepsilon  }_{ r } } \sum _{ n=-N }^{ N }{ { \left( -1 \right)  }^{ n } } \left| z-na \right|  \right\} \quad \quad \quad \label{C.40}
 \end{equation} 

The function \ref{C.40} can only be normalized when $\sigma \rightarrow { \left( -1 \right)  }^{ N }\sigma $, , it is $l\rightarrow l{ \left( -1 \right)  }^{ N }$, since it diverges when $z\rightarrow -\infty $, in this way to find the constant: $\Lambda $:

\renewcommand{\theequation}{\thesection.\arabic{equation}}
\begin{equation}
\int _{ 0 }^{ \infty  }{ { \left| \Psi \left( z \right)  \right|  }^{ 2 } } dz=1 \label{C.41}
\end{equation}

\renewcommand{\theequation}{\thesection.\arabic{equation}}
\begin{equation}
{ \left| \Lambda  \right|  }^{ 2 }\int _{ 0 }^{ \infty  }{ \text{exp}\left\{ \frac { -1 }{ l{ \varepsilon  }_{ r } } \sum _{ n=-N }^{ N }{ { \left( -1 \right)  }^{ n } } \left| z-na \right|  \right\} dz } =1 \label{C.42}
\end{equation}

Employing, 

\renewcommand{\theequation}{\thesection.\arabic{equation}}
\begin{equation}
\sum _{ n=-N }^{ N }{ { \left( -1 \right)  }^{ n } } \left| z-na \right| =\sum _{ n=-N }^{ 0 }{ { \left( -1 \right)  }^{ n } } \left| z-na \right| \quad +\sum _{ n=-1 }^{ N }{ { \left( -1 \right)  }^{ n } } \left| z-na \right| \quad \label{C.43}
\end{equation}

Changing $n$ by $-n$ in the first sum on the right:

\renewcommand{\theequation}{\thesection.\arabic{equation}}
\begin{equation}
\sum _{ n=-N }^{ N }{ { \left( -1 \right)  }^{ n } } \left| z-na \right| =\sum _{ -n=-N }^{ 0 }{ { \left( -1 \right)  }^{ -n } } \left| z+na \right| \quad +\sum _{ n=1 }^{ N }{ { \left( -1 \right)  }^{ n } } \left| z-na \right| \quad \label{C.44}
\end{equation}

\renewcommand{\theequation}{\thesection.\arabic{equation}}
\begin{equation}
\sum _{ n=-N }^{ N }{ { \left( -1 \right)  }^{ n } } \left| z-na \right| =\sum _{ n=0 }^{ N }{ { \left( -1 \right)  }^{ n } } \left| z+na \right| \quad +\sum _{ n=1 }^{ N }{ { \left( -1 \right)  }^{ n } } \left| z-na \right| \quad  \label{C.45}
\end{equation}

Replacing \ref{C.45} in \ref{C.42} the integral to be resolved is: 

\renewcommand{\theequation}{\thesection.\arabic{equation}}
\begin{equation}
\begin{matrix} \int _{ 0 }^{ Na }{ \text{exp}\left\{ \frac { -1 }{ l{ \varepsilon  }_{ r } } \left\{ \sum _{ n=0 }^{ N }{ { \left( -1 \right)  }^{ n } } \left| z+na \right| +\sum _{ n=1 }^{ N }{ { \left( -1 \right)  }^{ n } } \left| z-na \right|  \right\}  \right\} dz } + \\  \\ \int _{ Na }^{ \infty  }{ \text{exp}\left\{ \frac { -1 }{ l{ \varepsilon  }_{ r } } \left\{ \sum _{ n=0 }^{ N }{ { \left( -1 \right)  }^{ n } } \left| z+na \right| +\sum _{ n=1 }^{ N }{ { \left( -1 \right)  }^{ n } } \left| z-na \right|  \right\}  \right\} dz }  \end{matrix} \label{C.46}
\end{equation}

As,

\renewcommand{\theequation}{\thesection.\arabic{equation}}
\begin{equation}
\int _{ 0 }^{ \infty  }{ { e }^{ f\left( z \right)  }dz } =\int _{ 0 }^{ Na }{ { e }^{ f\left( z \right)  }dz } +\int _{ Na }^{ \infty  }{ { e }^{ f\left( z \right)  }dz } \\ \quad \label{C.47}
\end{equation}

Being  $f\left( z \right)$:

\renewcommand{\theequation}{\thesection.\arabic{equation}}
\begin{equation}
f\left( z \right) =\frac { -1 }{ l{ \varepsilon  }_{ r } } \left\{ \sum _{ n=0 }^{ N }{ { \left( -1 \right)  }^{ n } } \left| z+na \right| +\sum _{ n=1 }^{ N }{ { \left( -1 \right)  }^{ n } } \left| z-na \right|  \right\} \label{C.48}
\end{equation}

For $Na<z<\infty $ the factor  $f\left( z \right) $ is reduced to,

\renewcommand{\theequation}{\thesection.\arabic{equation}}
\begin{equation}
f\left( z \right) =\frac { -1 }{ l{ \varepsilon  }_{ r } } \left( z+\sum _{ n=1 }^{ N }{ { \left( -1 \right)  }^{ n }2z }  \right) \quad \quad \quad \quad \quad \\ \quad  \label{C.49}
\end{equation}

Solving the second integral of \ref{C.47} with the factor \ref{C.49}:

\renewcommand{\theequation}{\thesection.\arabic{equation}}
\begin{equation}
\int _{ Na }^{ \infty  }{ { e }^{ f\left( z \right)  } } dz=\int _{ Na }^{ \infty  }{ exp\left\{ \frac { -1 }{ l{ \varepsilon  }_{ r } } \left( z+\sum _{ n=1 }^{ N }{ { \left( -1 \right)  }^{ n }2z }  \right)  \right\} dz } \quad \quad \quad \quad \quad \\ \quad  \label{C.50}
\end{equation}

Expanding $\sum _{ n=1 }^{ N }{ { \left( -1 \right)  }^{ n }2z } =\left( -1+{ \left( -1 \right)  }^{ N } \right) z$,

\renewcommand{\theequation}{\thesection.\arabic{equation}}
\begin{equation}
\int _{ Na }^{ \infty  }{ { e }^{ f\left( z \right)  } } dz=\int _{ Na }^{ \infty  }{ exp\left\{ \frac { -z }{ l{ \varepsilon  }_{ r } } { \left( -1 \right)  }^{ N } \right\} dz } =l{ \varepsilon  }_{ r }{ \left( -1 \right)  }^{ N } \text{exp}\left.\left\{ \frac { -z }{ l{ \varepsilon  }_{ r } } { \left( -1 \right)  }^{ N } \right\}\right|_{Na}^\infty  \label{C.51}
\end{equation}

This integral will diverge unless $l=l{ \left( -1 \right)  }^{ N }, { l }^{ -1 }={ l }^{ -1 }{ \left( -1 \right)  }^{ N }$, therefore: 

\renewcommand{\theequation}{\thesection.\arabic{equation}}
\begin{equation}
\int _{ Na }^{ \infty  }{ { e }^{ f\left( z \right)  } } dz=l{ \varepsilon  }_{ r } \text{exp}\left( \frac { -Na }{ l{ \varepsilon  }_{ r } }  \right) \label{C.52}
\end{equation}

For the limit  $ka<z<\left( k+1 \right) a$ the factor  $f\left( z \right) $ is reduced to,

\renewcommand{\theequation}{\thesection.\arabic{equation}}
\begin{equation}
f\left( z \right) =\frac { -1 }{ l{ \varepsilon  }_{ r } } \left( { \left( -1 \right)  }^{ k }z+\sum _{ n=k+1 }^{ N }{ { \left( -1 \right)  }^{ n }2na }  \right) \label{C.53}
\end{equation}

Substituting \ref{C.53} in \ref{C.47} to solve the first integral 

\renewcommand{\theequation}{\thesection.\arabic{equation}}
\begin{equation}
\int _{ 0 }^{ Na }{ { e }^{ f\left( z \right)  } } dz=\sum _{ k=0 }^{ N-1 }{ \int _{ ka }^{ \left( k+1 \right) a }{ \text{exp}\left\{ \frac { -1 }{ l{ \varepsilon  }_{ r } } \left( { \left( -1 \right)  }^{ k }z+\sum _{ n=k+1 }^{ N }{ { \left( -1 \right)  }^{ n }2na }  \right)  \right\}  } dz } \label{C.54}
\end{equation}

Expanding the exponential,

\renewcommand{\theequation}{\thesection.\arabic{equation}}
\begin{equation}
\quad \quad \begin{matrix} \int _{ 0 }^{ Na }{ { e }^{ f\left( z \right)  } } dz=l{ \varepsilon  }_{ r }\sum _{ k=0 }^{ N-1 }{ { \left( -1 \right)  }^{ k } \text{exp}\left\{ \frac { -1 }{ l{ \varepsilon  }_{ r } } \left( \sum _{ n=k+1 }^{ N }{ { \left( -1 \right)  }^{ n }2na }  \right)  \right\}  }  \\  \\ \text{exp}\left\{ \frac { -{ \left( -1 \right)  }^{ k }ka }{ l{ \varepsilon  }_{ r } }  \right\} \left[ 1-\text{exp}\left\{ \frac { -{ \left( -1 \right)  }^{ k }ka }{ l{ \varepsilon  }_{ r } }  \right\}  \right]  \end{matrix} \label{C.55}
\end{equation}

Simplifying terms and assuming $2\sinh { \left( \frac { x }{ 2 }  \right)  } ={ e }^{ \frac { x }{ 2 }  }\left( 1-{ e }^{ -x } \right) $, it is found that: 

\renewcommand{\theequation}{\thesection.\arabic{equation}}
\begin{equation}
\int _{ 0 }^{ Na }{ { e }^{ f\left( z \right)  } } dz=2l{ \varepsilon  }_{ r }N\text{exp}\left\{ \frac { -\left( 1+2N \right) a }{ 2l{ \varepsilon  }_{ r } }  \right\} \sinh { \left( \frac { a }{ 2l{ \varepsilon  }_{ r } }  \right)  } \label{C.56}
 \end{equation}

Solved the two integrals of \ref{C.47} it is obtained: 

\renewcommand{\theequation}{\thesection.\arabic{equation}}
\begin{equation}
\int _{ 0 }^{ \infty  }{ { e }^{ f\left( z \right)  } } dz=l{ \varepsilon  }_{ r } \text{exp}\left\{ \frac { -Na }{ l{ \varepsilon  }_{ r } }  \right\} +2l{ \varepsilon  }_{ r }N \text{exp}\left\{ \frac { -\left( 1+2N \right) a }{ 2l{ \varepsilon  }_{ r } }  \right\} \sinh { \left( \frac { a }{ 2l{ \varepsilon  }_{ r } }  \right)  } \quad \quad \label{C.57}
\end{equation}

Consequently, the normalization constant turns out to be: 

\renewcommand{\theequation}{\thesection.\arabic{equation}}
\begin{equation}
{ \Lambda  }^{ -1 }={ \left( l{ \varepsilon  }_{ r } \text{exp}\left\{ \frac { -Na }{ l{ \varepsilon  }_{ r } }  \right\} +2l{ \varepsilon  }_{ r }N \text{exp}\left\{ \frac { -\left( 1+2N \right) a }{ 2l{ \varepsilon  }_{ r } }  \right\} \sinh { \left( \frac { a }{ 2l{ \varepsilon  }_{ r } }  \right)  }  \right)  }^{ \frac { 1 }{ 2 }  }\quad \quad \label{C.58}
\end{equation}

In this way, the eigenfunction  \ref{C.40} is rewritten by replacing the normalization constant with the expression found in \ref{C.58}:

\renewcommand{\theequation}{\thesection.\arabic{equation}}
\begin{equation}
\Psi \left( z \right) =\quad \frac { \prod _{ n=-N }^{ N }{ \text{exp}\left\{ \frac { -{ \left( -1 \right)  }^{ n+N }\left| z-na \right|  }{ 2l{ \varepsilon  }_{ r } }  \right\}  }  }{ { \left( l{ \varepsilon  }_{ r } \text{exp}\left\{ \frac { -Na }{ l{ \varepsilon  }_{ r } }  \right\} +2l{ \varepsilon  }_{ r }N \text{exp}\left\{ \frac { -\left( 1+2N \right) a }{ 2l{ \varepsilon  }_{ r } }  \right\} \sinh { \left( \frac { a }{ 2l{ \varepsilon  }_{ r } }  \right)  }  \right)  }^{ \frac { 1 }{ 2 }  } } \quad \label{C.59}
\end{equation}

Therefore, if we make the appropriate substitutions ${ l }^{ -1 }=\frac { 2m{ \alpha  }_{ 1 } }{ { \hslash  }^{ 2 } } $ , we will find an eigenfunction that models its own state of bound energy, corresponding to a quantum particle of mass m that moves in a potential field, similar to that of a one-dimensional ionic crystal, presenting a linear dielectric constant for a given frequency range \cite{H. Sadri}: 

\renewcommand{\theequation}{\thesection.\arabic{equation}}
\begin{equation}
\Psi \left( z \right) =\quad \frac { \prod _{ n=-N }^{ N }{ \text{exp}\left\{ \frac { -{ \alpha  }_{ 1 }m{ \left( -1 \right)  }^{ n+N }\left| z-na \right|  }{ { \hslash  }^{ 2 }{ \varepsilon  }_{ r } }  \right\}  }  }{ { \left( { \varepsilon  }_{ r }\frac { { \hslash  }^{ 2 } }{ { 2\alpha  }_{ 1 }m } \text{exp}\left\{ \frac { -2{ \alpha  }_{ 1 }mNa }{ { \hslash  }^{ 2 }{ \varepsilon  }_{ r } }  \right\} +{ \varepsilon  }_{ r }\frac { { \hslash  }^{ 2 } }{ m{ \alpha  }_{ 1 } } N \text{exp}\left\{ \frac { -{ \alpha  }_{ 1 }m\left( 1+2N \right) a }{ { \hslash  }^{ 2 }{ \varepsilon  }_{ r } }  \right\} \sinh { \left( \frac { am{ \alpha  }_{ 1 } }{ { \hslash  }^{ 2 }{ \varepsilon  }_{ r } }  \right)  }  \right)  }^{ \frac { 1 }{ 2 }  } } \quad  \label{C.60}
\end{equation}

\section{Derivation of the eigenfunction (16)}
\label{sec:D}

Considering the general case, in a distribution of parallel plates with $N\ge 1$, and surface current densities ${ k }_{ 0 }=\sum _{ -N }^{ N }{ { \left( -1 \right)  }^{ n } } { k }_{ 0 }$, the magnetic vector potential is determined in a dimension such as:  

\renewcommand{\theequation}{\thesection.\arabic{equation}}
\begin{equation}
{ A }_{ x }\left( z \right) =\frac { -{ \mu  }{ k }_{ 0 } }{ 2 } \sum _{ n=-N }^{ N }{ { \left( -1 \right)  }^{ n+N } } \left| z-nb \right| \label{D.61}
\end{equation}

Substituting the magnetic potential of \ref{D.61} in the eigenfunctiona Eq. \ref{Eq.11} and performing a normalization to find the constant $C$:

\renewcommand{\theequation}{\thesection.\arabic{equation}}
\begin{equation}
\int _{ 0 }^{ \infty  }{ { \left| \Psi \left( z \right)  \right|  }^{ 2 } } dz=1 \label{D.62}
\end{equation}

Making appropriate substitutions and being  $L$ an arbitrary length: 

\renewcommand{\theequation}{\thesection.\arabic{equation}}
\begin{equation}
\left[ \frac { { \mu  }_{ 0 }{ k }_{ 0 } }{ { A }_{ 0 } }  \right] =\left[ \frac { { A }_{ 0 }m{ b }_{ 0 }^{ 3 }{ k }_{ 0 } }{ { \hslash  }^{ 2 } }  \right] =\left[ \frac { 1 }{ L }  \right] \label{D.63}
\end{equation}

\renewcommand{\theequation}{\thesection.\arabic{equation}}
\begin{equation}
{ \left| C \right|  }^{ 2 }\int _{ 0 }^{ \infty  }{ \text{exp}\left\{ \frac { -{ \mu  }_{ r } }{ L } \sum _{ n=-N }^{ N }{ { \left( -1 \right)  }^{ n } } \left| z-nb \right|  \right\} dz } =1\quad \quad \label{D.64}
\end{equation}

Using a procedure analogous to that established in the appendix to find the normalization constant, we obtain:

\renewcommand{\theequation}{\thesection.\arabic{equation}}
\begin{equation}
{ C }^{ -1 }={ \left( \frac { L }{ { \mu  }_{ r } } \text{exp}\left\{ \frac { -{ \mu  }_{ r }Nb }{ L }  \right\} +2\frac { L }{ { \mu  }_{ r } } N \text{exp}\left\{ \frac { -{ \mu  }_{ r }\left( 1+2N \right) b }{ 2L }  \right\} \sinh { \left( \frac { { \mu  }_{ r }b }{ 2L }  \right)  }  \right)  }^{ \frac { 1 }{ 2 }  }\quad \quad \label{D.65}
\end{equation}

By replacing the normalization constant \ref{D.65} and the magnetic potential Eq. \ref{D.62} in \ref{Eq.11} an expression similar to \ref{C.60}: 

\renewcommand{\theequation}{\thesection.\arabic{equation}}
\begin{equation}
\quad \Psi \left( z \right) =\quad \frac { \prod _{ n=-N }^{ N }{ \text{exp}\left\{ \frac { -{ \mu  }_{ r }{ \left( -1 \right)  }^{ n+N }\left| z-nb \right|  }{ 2L }  \right\}  }  }{ { \left( \frac { L }{ { \mu  }_{ r } } \text{exp}\left\{ \frac { -{ \mu  }_{ r }Nb }{ L }  \right\} +2\frac { L }{ { \mu  }_{ r } } N \text{exp}\left\{ \frac { -{ \mu  }_{ r }\left( 1+2N \right) b }{ 2L }  \right\} \sinh { \left( \frac { { \mu  }_{ r }b }{ 2L }  \right)  }  \right)  }^{ \frac { 1 }{ 2 }  }\quad  } \quad  \label{D.66}
\end{equation}

By developing appropriate substitutions  ${ L }^{ -1 }=\frac { 2{ \alpha  }_{ 2 }m }{ { \hslash  }^{ 2 } } $  in \ref{D.66}, it is found that it is also possible to model an eigenfunction that describes the behavior of a quantum particle of mass m that crosses a one-dimensional ionic crystal with potential energy, even if it has a linear magnetic constant; in this way, \ref{D.66} is expressed in a general way by means of Eq. \ref{Eq.17}.

\end{appendices}

%

\begin{thebibliography}{}



%

\bibitem{De la Pena} L. De la Peña, \textit{Introduccion a la Mecanica Cuantica}. {Fondo de Cultura Economica, Mexico, Ciudad de Mexico, 2012}. 

\bibitem{P. Weinberger} P. Weinberger, ``Revisiting Louis de Broglie's famous 1924 paper in the Philosophical Magazine", Philosophical Magazine Letters.,  Vol. 86, no. 7, pp.405-410, 2006. \href{https://doi.org/10.1080/09500830600876565}{doi:10.1080/09500830600876565}  


\bibitem{E. Schrodinger} E. Schrödinger, ``An Undulatory Theory of the Mechanics of Atoms and Molecules", Physical Review., Vol. 28, no. 1049, 1926. 

\bibitem{A. Dirac} P. A. Dirac, ``The quantum theory of the electron, The Royal Society., Vol. 117, no. 610, 1928. 


\bibitem{G. Gonzalez} G. Gonzalez, ``Relation between Poisson and Schrödinger equations", Am. J. Phys., Vol. 80, no. 715, 2012. \href{https://doi.org/10.1119/1.4722788}{doi:10.1119/1.4722788}

\bibitem{V. Rokaj} V. Rokaj, F. k. Diakonos and G. Gabriel, ``Comment on and Erratum: “Relation between Poisson and Schrödinger equations”", Am. J. Phys., Vol. 82, no. 802, 2014. \href{https://doi.org/10.1119/1.4884037}{doi:10.1119/1.4884037}

\bibitem{N. Ben Abdallah} N. Ben Abdallah, P. Degond, P. A, ``On a one-dimensional Schrödinger-Poisson scattering model", Math. Phys., Vol. 48, no. 135, 1997. \href{https://doi.org/10.1007/PL00001463}{doi:10.1007/PL00001463}

\bibitem{G. Gonzalez2} G. Gonzalez, ``Erratum: ``Relation between Poisson and Schrödinger equations" (Am. J. Phys. 80, 715 ( 2012))", Am. J. Phys., Vol. 80, no. 12, 2012. \href{https://doi.org/10.1119/1.4765458}{doi:10.1119/1.4765458}

\bibitem{M. Mohammad-Ali} M. Mohammad-Ali et al, ``Supersymmetric Optical Structures", Phys. Rev. Lett., Vol. 110, no. 23, 2013. \href{https://doi.org/10.1103/PhysRevLett.110.233902}{doi:10.1103/PhysRevLett.110.233902}

\bibitem{W. Tan} W. Tan et al, ``Photonic simulation of topological excitations in metamaterials", Sci. Rep., Vol. 4, no. 3842, 2015. \href{https://doi.org/10.1038/srep03842}{doi:10.1038/srep03842}

\bibitem{L. Lamata} L. Lamata et al, ``Dirac Equation and Quantum Relativistic Effects in a Single Trapped Ion", Phys. Rev. Lett., Vol. 98, no. 253005, 2007. \href{https://doi.org/10.1103/PhysRevLett.98.253005}{doi:10.1103/PhysRevLett.98.253005}


\bibitem{A. Arnold} A. Arnold, ``Mathematical concepts of open quantum boundary conditions", Transp. Theory Stat. Phys., Vol. 30, no. 561, 2006. \href{https://doi.org/10.1081/TT-100105939}{doi:10.1081/TT-100105939}

\bibitem{I. Richard} I. Richard Lapidus, ``One‐dimensional models for two‐electron systems", Am. J. Phys., Vol. 43, no. 790, 1975. \href{https://doi.org/10.1119/1.9708}{doi:10.1119/1.9708}

\bibitem{M. Born} M. Born, K. Huang, \textit{Dynamical Theory of Crystal Lattices}, {Oxford University Press, EE.UU., New York, 1998}.

\bibitem{H. Sadri} H. Sadri, \textit{Mathematical Methods: For Students of Physics and Related Fields},  {Springer, EE.UU., New York, 2008}.


\bibitem{A. Rabinovitch} A. Rabinovitch, ``Negative energy states of an ‘‘inverted’’ delta potential: Influence of boundary conditions", Am. J. Phys., Vol. 53, no. 768, 1985. \href{https://doi.org/10.1119/1.14310}{doi:10.1119/1.14310}

\bibitem{A. Stratton} A. Stratton, \textit{Electromagnetic Theory}, {John Wiley and Sons, EE. UU., New Jersey, 2007}.

\bibitem{I. E. Irodov} I. E. Irodov, \textit{Basic laws of electromagnetism}, {CBS Publishers and Distributors, India, New Delhi, 1994}.

\bibitem{A. Sepulveda} A. Sepulveda Soto, \textit{Electromagnetismo}, {Universidad de Antioquia, Colombia, Antioquia, 2009}.

\bibitem{F.J. Dyson} F.J. Dyson, ``Ground‐State Energy of a Finite System of Charged Particles", Am. J. Phys., Vol. 8, no. 1538, 1967. \href{https://doi.org/10.1063/1.1705389}{doi:10.1063/1.1705389}

\bibitem{J. D. Griffiths} J. D. Griffiths, \textit{Introduction to Electrodynamics}, {Cambridge University Press, EE.UU., New York, 2017}.

\bibitem{F. C. Brown} F. C. Brown,  \textit{Propiedades dieléctricas y ópticas de los cristales iónicos}, {Reverte, España, Barcelona, 1970}.



\end{thebibliography}
%

\end{document}